\documentclass[twocolumn,aps,pra,10pt,superscriptaddress]{revtex4-1}

\usepackage{epsfig}
\usepackage{amsmath}
\usepackage{amssymb}

\newcommand{\ket}[1]{|#1\rangle}

\begin{document}

\title{Non-classicality thresholds for multiqubit states - numerical analysis}

\author{Jacek Gruca}
\affiliation{Institute of Theoretical Physics and Astrophysics, University of Gda\'nsk, PL-80-952 Gda\'nsk, Poland}

\author{Wies{\l}aw Laskowski}
\email{wieslaw.laskowski@univ.gda.pl}
\affiliation{Institute of Theoretical Physics and Astrophysics, University of Gda\'nsk, PL-80-952 Gda\'nsk, Poland}
\affiliation{Fakult\"at f\"ur Physik, Ludwig-Maximilians Universit\"at M\"unchen, D-80799 M\"unchen, Germany}
\affiliation{Max-Planck Institut f\"ur Quantenoptik, D-85748 Garching, Germany}

\author{Marek \.Zukowski}
\affiliation{Institute of Theoretical Physics and Astrophysics, University of Gda\'nsk, PL-80-952 Gda\'nsk, Poland}

\author{Nikolai Kiesel}
\affiliation{Faculty of Physics, University of Vienna, A-1090 Vienna, Austria}

\author{Witlef Wieczorek}
\altaffiliation[Present address: ]{Faculty of Physics, University of Vienna,  
Boltzmanngasse 5, A-1090 Wien, Austria}
\affiliation{Fakult\"at f\"ur Physik, Ludwig-Maximilians Universit\"at M\"unchen, D-80799 M\"unchen, Germany}
\affiliation{Max-Planck Institut f\"ur Quantenoptik, D-85748 Garching, Germany}

\author{Christian Schmid}
\affiliation{Fakult\"at f\"ur Physik, Ludwig-Maximilians Universit\"at M\"unchen, D-80799 M\"unchen, Germany}
\affiliation{Max-Planck Institut f\"ur Quantenoptik, D-85748 Garching, Germany}
\affiliation{European Organisation for Astronomical Research in the Southern Hemisphere, D-85748 Garching, Germany}

\author{Harald Weinfurter}
\affiliation{Fakult\"at f\"ur Physik, Ludwig-Maximilians Universit\"at M\"unchen, D-80799 M\"unchen, Germany}
\affiliation{Max-Planck Institut f\"ur Quantenoptik, D-85748 Garching, Germany}

\date{\today}

\begin{abstract}
States that strongly violate Bell's inequalities are required in many quantum-informational protocols as, for example, in cryptography, secret sharing, and the reduction of communication complexity. We investigate families of such states with a numerical method which allows us to reveal non-classicality even without direct knowledge of Bell's inequalities for the given problem. An extensive set of numerical results is presented and discussed.
\end{abstract}

\maketitle

\section{Introduction}

Until the discovery of the protocol of entanglement-based quantum cryptography \cite{Ekert91}, Bell's theorem was relevant only in discussions concerning foundations of quantum mechanics. It established means to evaluate the most reasonable class of hidden variable theories, i.e., local and realistic theories. Once the theorem proved useful in quantum cryptography and other quantum communication schemes (secret sharing \cite{ZukEtAll1998ActaPhysPolA}, reduction of communication complexity \cite{Burhman1997, Galvao2002, BrukZuk2004}, etc.), it gained an additional technical status. Today, we are not only interested in showing that given quantum correlations do not admit a local hidden variable model, but also, with Bell inequalities, we can now prove the usefulness of states in quantum protocols. The strength of such violations indicates resistance to noise or to decoherence. 

However, only in simple cases, like two dichotomic measurements per observer, one can pinpoint the full set of such Bell-type inequalities, and put them into a manageable form of just one single non-linear inequality \cite{WZ,WW}. In other cases we know usually only some subsets of tight inequalities for the given problem. These subsets still contain a vast number of different inequalities.

One of the methods to avoid the problem of both missing inequalities and handling the exploding numbers of known ones is to address the problem numerically, without a direct use of Bell's inequalities \cite{ZukowskiBaturo98}. It is very easy to notice that the problem of existence of a local hidden variable model for the given set of quantum probabilities is a form of a linear programming problem (see Sec. \ref{LP}). Thanks to such an approach it was possible to show for example that non-classicality of two-qudit correlations is higher than for qubits, and that it increases with the dimension \cite{KASZLIKOWSKI-SR}. 

Below we present how linear programming can be harnessed to study non-classicality of quantum correlations. This approach was put in the form of a computer code: \textsc{Steam-roller}. We will present the numerical results in Secs. III and IV. Here we particularly aimed at multi-setting scenarios for multi-qubit states which are known to possess interesting kinds of non-classical behavior. Most of our results cover regions for which the full set of Bell's inequalities is unknown. Even the results for two settings per observer cover a new territory: they are produced for the full set of probabilities for the given $N$-qubit problem. In the case of the WWWZB inequalities \cite{WZ,WW,ZB} only highest order correlation functions \cite{n1} are taken into account. These do not reflect interference effects that may occur between subsets of particles. Thus the WWWZB set, despite being complete for highest order correlation functions (in the case of a two setting choice per observer), is not complete for the full description of quantum phenomena. This full set of probabilities for the given problem gives full knowledge about an experiment, and therefore contains all available information. 

\section{Linear programming}
\label{LP}

In the most general Bell's experiment, $N$ observers perform measurements on a given state $\rho_N$. Each observer can choose between $m_i$ arbitrary dichotomic observables ($i=1,\dots,N$). Their measurement results will exhibit correlations, not describable by classical statistics (i.e., by classical realistic models). For brevity let us consider a case with two observers (Alice and Bob) and two measurement settings per side. In this case the local realistic models are equivalent to the existence of a joint probability distribution $p_{lr}(a_1,a_2,b_1,b_2)$, where $a_i =\pm 1$ denotes the result of the measurement of Alice's $i$-th observable (Bob's results are denoted by $b_k = \pm 1$). Quantum predictions for the probabilities are the marginal sums of the above joint probability distribution $p_{lr}$
\begin{align}
\label{marginals}
P_{QM}(r_a, r_b| A_1, B_1) &=  \sum_{a_2, b_2 = \pm1} p_{lr}(r_a, a_2, r_b, b_2)\nonumber\\
P_{QM}(r_a, r_b| A_1, B_2) &=  \sum_{a_2, b_1 = \pm1} p_{lr}(r_a, a_2, b_1, r_b)\nonumber\\
P_{QM}(r_a, r_b| A_2, B_1) &=  \sum_{a_1, b_2 = \pm1} p_{lr}(a_1, r_a, r_b, b_2)\nonumber\\
P_{QM}(r_a, r_b| A_2, B_2) &=  \sum_{a_1, b_1 = \pm1} p_{lr}(a_1, r_a, b_1, r_b),
\end{align}
where $P_{QM}(r_a, r_b| A_i, B_k)$ denotes the probability of obtaining the result $r_a$ by Alice and $r_b$ by Bob, when they measure the observables $A_i$ and $B_k$, respectively.  It has been shown that for some entangled states, no local realistic probability distribution $p_{lr}(a_1, a_2, b_1, b_2)$ exists, which could satisfy the set of equalities (\ref{marginals}). Therefore, no local realistic model can reproduce the predictions of quantum mechanics for these entangled states. This statement is known as Bell's theorem \cite{BELL}.

This is one of the ways one can express Bell's theorem. The more frequent approach is to find certain (Bell) inequalities which are satisfied by local realistic models, but violated by some quantum predictions. More importantly, there are many statements on the ``strength'' of a violation of such inequalities which use such terms as ``amount of violation'' or ``factor of violation.'' However, such measures cannot be used for direct comparison between different Bell inequalities or with a different method, such as the one used in our paper. Thus, following \cite{KASZLIKOWSKI-SR}, we shall use a ``violation'' parameter $v_{crit}$ which seems to be much more objective: we shall seek the amount of random (``white'') noise admixture that is required to completely hide the 
 non-classical
character of the original correlations for the given state.

If we mix some amount of white noise to the two-qubit state $\rho$, we obtain a state described by the following density operator:
\begin{equation}
\rho(v) = v \rho + \frac{1-v}{2^2} \openone^{\otimes 2}.
\end{equation}
The quantum probability $P^{v}_{QM}(r_a, r_b| A_i, B_k) \equiv P(r_a, r_b| A_i, B_k)$ for the state $\rho(v)$ reads 
\begin{equation}
P(r_a, r_b | A_i, B_k) = v P_{QM}(r_a, r_b | A_i, B_k) + \frac{1}{2^2}(1-v).
\label{probability}
\end{equation}
The parameter $v$ is the visibility of the state, and obviously $(1-v)$ is the amount of noise admixture. For $v=0$ the equalities (\ref{marginals}) are obviously satisfied (white noise has a local realistic model). For $v=1$ the equalities (\ref{marginals}) can not be satisfied for some entangled states $\rho$ (and for a particular choice of observables). For such states there exists the critical visibility $v_{crit}$ that for $v\leqslant v_{crit}$ there exists a local realistic probability distribution $p_{lr}(a_1, a_2, b_1, b_2)$ that satisfies the set of equalities (\ref{marginals}).

A set of $16$ probabilities $P_{QM}(r_a, r_b | A_i, B_k)$ corresponds to a single point in a $16$-dimensional space. Those sets which can be described using local realistic models form a convex geometric figure called a Bell-Pitovsky polytope in this $16$-dimensional space. The facets of the polytope are equivalent to tight Bell inequalities. For $v = v_{crit}$ the local realistic probabilities $p_{lr}(a_1, a_2, b_1, b_2)$ lead to a set of probabilities $P_{QM}(r_a, r_b | A_i, B_k)$ which form the coordinates of  a geometric point which lies in a facet of the polytope (i.e., they saturate a Bell inequality). It has to be noted, however, that our numerical method does not lead to uncovering particular Bell inequalities which are violated by an investigated state. We think that, the power of the method lies in the fact that one does not need any knowledge at all of the forms of Bell inequalities when analyzing non-classicality of quantum-entangled states. Seeking violated Bell inequalities is highly inefficient, since even in the simplest cases one ends up with an exploding number of Bell inequalities, the bulk of them trivial. We skip this cumbersome step entirely, see the following paragraphs.

As mentioned, critical visibility $v_{crit}$ depends on the particular set of observables Alice and Bob choose from. We can parametrize any dichotomic observable $X_i$ by the two angles $\theta^X_i$ and $\phi^X_i$ ($X=A,B$) in the following way:
\begin{align}
X &= |+\rangle_X\langle +| - |-\rangle_X\langle -|, \nonumber\\
|{\pm}\rangle_X &= \cos(\pm \pi/4 + \theta^X_i)|0\rangle_X +  e^{i\phi^X_i}\sin(\pm \pi/4+\theta^X_i)|1\rangle_X.
\end{align}
Now, the quantum probability can be calculated: $P_{QM}(r_a, r_b | A_i, B_k) = {\rm Tr}(\rho |r_a\rangle_A\langle r_a| \otimes |r_b\rangle_B\langle r_b|)$.
Hence we can define the critical visibility function of the angles
\begin{align}
\label{v_function}
v_{crit}(\theta^A_1, \phi^A_1, \theta^A_2, \phi^A_2, \theta^B_1, \phi^B_1, \theta^B_2, \phi^B_2) \in [0,1].
\end{align}

Our task is to find, for a given state $\rho$, the minimal critical visibility. 
%
%
If the two observers $A$ and $B$ perform two measurements each, the probabilities (\ref{probability}) are then parametrized by the angles:
\begin{align}
P(r_a, r_b | A_i, B_k) \equiv P(r_a, r_b, \theta^A_i, \phi^A_i, \theta^B_k, \phi^B_k).
\end{align}
The critical visibility is obtained by maximization of the visibility until the set of equalities (\ref{marginals}) can no longer be satisfied. This is done by means of linear programming. We adopt
\begin{align}
p_0 &\equiv p_{lr}(-, \ldots, -, -),\nonumber\\
p_1 &\equiv p_{lr}(-, \ldots, -, +),\nonumber\\
&\ldots\nonumber\\
p_{n-1} &\equiv p_{lr}(+, \ldots, +, +),
\end{align}
where $n=2^{m_1+\ldots+m_N}$ and hereby give a complete description of the linear programming problem to be solved (based on Eqns. (\ref{marginals}) and (\ref{probability})) for the case with Alice and Bob:
\begin{widetext}
\begin{eqnarray}
\label{problem}
&&\sum_{a_2, b_2 = \pm1} p_{lr}(r_a, a_2, r_b, b_2) = \frac{1-v}{2^2} + v P(r_a, r_b, \theta^A_1, \phi^A_1, \theta^B_1, \phi^B_1),\nonumber\\
&&\sum_{a_2, b_1 = \pm1} p_{lr}(r_a, a_2, b_1, r_b) = \frac{1-v}{2^2} + v P(r_a, r_b, \theta^A_1, \phi^A_1, \theta^B_2, \phi^B_2),\nonumber\\
&&\sum_{a_1, b_2 = \pm1} p_{lr}(a_1, r_a, r_b, b_2) = \frac{1-v}{2^2} + v P(r_a, r_b, \theta^A_2, \phi^A_2, \theta^B_1, \phi^B_1), \nonumber\\
&&\sum_{a_1, b_1 = \pm1} p_{lr}(a_1, r_a, b_1, r_b) = \frac{1-v}{2^2} + v P(r_a, r_b, \theta^A_2, \phi^A_2, \theta^B_2, \phi^B_2), \nonumber\\
&&\sum_{k=0}^{n-1} p_k =1; ~~0\leqslant v \leqslant1; ~~ 0\leqslant p_k \leqslant1, \nonumber\\
&&z(p_0, p_1, \ldots, p_{n-1}, v) = v,
\end{eqnarray}
\end{widetext}
where $z$ is the (trivial) function to be maximized. 

In the general case the four expressions above evaluate to $c = 2^N m_1 \cdots m_N$ equalities (e.g. for 3 qubits and 6 measurement settings per side, there are 1728 equalities). The above is known as the canonical form of the linear programming problem. We use the revised simplex method \cite{simplex} from the {\ttfamily GNU Linear Programming Kit} \cite{GLPK} to solve this problem. The top constraints form a simplex in the \mbox{($n+1$)-dimensional} space spanned by $p_0,\ldots,p_{n-1}$ and $v$. The algorithm traverses the vertices of the simplex until it finds the optimal solution. The returned solution is the global maximum of $z$ -- the critical visibility we seek. The minimization over the angles $\theta^A_1, \phi^A_1, \theta^B_1, \phi^B_1,\theta^A_2, \phi^A_2, \theta^B_2, \phi^B_2$ is realized by the downhill simplex method \cite{dsm} from the SciPy package \cite{scipy}.

Please note that the complexity of the linear programming problem is
defined by $n$ and $c$ -- values which increase exponentially with the
number of observers $N$ or the number of settings $m_i$ per observer. We
are, therefore, limited to the computational capabilities of today's
numerical machines when solving this problem. We used a machine
with an {\ttfamily Intel Pentium D CPU 3.20GHz} processor for our computations. Some of the results
in this paper took several weeks to compute. Computations of results
for bigger problems can be anticipated to complete in a reasonable time
when using faster machines in the future.

One might wonder about the credibility of the results presented in
this paper. There are two aspects which raise doubts, namely, numerical
precision and local minimum risk when computing the minimum of the
critical visibility function. Please note that there is no local
maximum risk when computing the value of the critical visibility
function - the simplex algorithm guarantees it will always find the
global maximum as described in \cite{simplex}. As for the precision of
the results, we will not conduct an extensive analysis but will
plainly state that this is not an issue. For all the results which can
 be calculated analytically, \textsc{Steam-roller} generated results are compliant on
all 14 displayed significant digits. For example, for a two-qubit Greenberger-Horne-Zeilinger
(GHZ) state, the analytically-derived minimal critical visibility is
 $\frac{\sqrt{2}}{2}$ and \textsc{Steam-roller} gives $0.70710678118655$. One of
 the reasons for such outstanding precision is that the numerical errors
do not propagate between consecutive critical visibility function
invocations. Another is that the linear programming library used
 properly handles numerical instabilities. The local minimum risk when
computing the critical visibility function is more of an issue. There
is no guarantee we have reached the global minimum, and \textsc{Steam-roller}
did fall into a local minimum from time to time. One should,  therefore, treat all the results in this paper in the following way;
 it is guaranteed that a given state has a minimal critical visibility
not greater than the one presented. As for the likelihood that it is
 also not smaller, we can only state that there's a good chance it is because \textsc{Steam-roller} provided us with a correct result in 99 cases out of
 100.

\section{Bounds on local realism}

We applied the numerical method to different types of quantum states, which are often discussed in the context of quantum information studies, namely,
\begin{itemize}
\item the generalized GHZ state \cite{GHZ,SCARANI}:
\begin{equation}\ket{\textrm{GHZ}(\alpha)}_N = \cos{\alpha} \ket{0 \cdots 0}_{N} + \sin{\alpha} \ket{1 \cdots 1}_{N},\nonumber \end{equation} 
\item the W state \cite{W}:
\begin{equation}\ket{\textrm{W}}_N = \frac{1}{\sqrt{N}}(\ket{10 \cdots 0}_{N} + \ket{010 \cdots 0}_{N} +\cdots + \ket{0 \cdots 01}_{N}), \nonumber \end{equation}
\item the four-qubit singlet state \cite{WZ, CABELLO}: 
\begin{eqnarray}\ket{\Psi_4} &=& \frac{1}{\sqrt{3}}(\ket{0011}+ \ket{1100})\nonumber \\ &-& \frac{1}{\sqrt{12}}(\ket{0101}+\ket{0110}+\ket{1001}+\ket{1010}),\nonumber \end{eqnarray}
\item the six-qubit  singlet state \cite{PSI6, CABELLO}:
$$\ket{\Psi_6} = \frac{1}{\sqrt{2}}\ket{GHZ^-}_6 + \frac{1}{2}(\ket{\overline{W}}_3\ket{W}_3
-\ket{W}_3\ket{\overline{W}}_3),$$ where $\ket{GHZ^-}_6=\frac{1}{\sqrt{2}}(\ket{000111}-\ket{111000})$,
  and  $\ket{\overline{W}}_3$ is the spin-flipped  $\ket{W}_3$,
\item the symmetric Dicke state \cite{DICKE}: 
$$\ket{D_N^{(N/2)}} = {N \choose N/2}^{-1/2}\sum_{\rm permutations} \ket{0 \cdots 0 \underbrace{1 \cdots 1}_{N/2} 0 \cdots 0}_N,$$
\item the four-qubit cluster state \cite{CLUSTER}:
$$\ket{\textrm{Cluster}} = \frac{1}{2}(\ket{0000} + \ket{0011} + \ket{1100} - \ket{1111}).$$
\end{itemize}

We calculated the critical visibility for an increasing number of different settings per side. 
All results are presented in Tab. \ref{tab-states}. They lead to the following observations. 

\begin{table}
\begin{tabular}{l l l l c} \hline \hline
No. & $N$ & State & $m_1~ \textrm{x}~ m_2 ~\textrm{x}~ \cdots ~\textrm{x}~ m_N$ & $v_{crit}$  \\ \hline \hline
\#1 & 2& $\ket{\textrm{GHZ}(\pi/4)}_2$   &  2x2 - 10x10    &     0.7071  \\ \hline 
\#2 & 3& $\ket{\textrm{GHZ}(\pi/4)}_3$   &  2x2x2 - 5x5x5  &     0.5000  \\ \cline{3-5}
\#3 &  & $\ket{\textrm{GHZ}(\pi/12)}_3$  &  2x2x2 - 4x4x4  &     0.8165  \\ \cline{3-5}
\#4 &  & $\ket{\textrm{GHZ}(\pi/180)}_3$ &  2x2x2 - 3x3x3  &     0.9988  \\ \cline{3-5}
\#5 &  & $\ket{W}_3$                     &  2x2x2 &  0.6442           \\ 
\#6 &  &                                 &  ~~{\footnotesize 3x3x2} &   ~~{\footnotesize 0.6330}  \\ 
\#7 &  &                                 &  ~~{\footnotesize 4x4x2} &   ~~{\footnotesize 0.6240}  \\ 
\#8 &  &                                 &  ~~{\footnotesize 5x5x2} &   ~~{\footnotesize 0.6236}  \\
\#9 &  &                                 &  ~~{\footnotesize 6x6x2} &   ~~{\footnotesize 0.6236}  \\ 
\#10&  &                                 &  3x3x3 &  0.6048           \\ 
\#11&  &                                 &  4x4x4 &  0.6013           \\ 
\#12&  &                                 &  5x5x5 &  0.6007           \\ \hline 
\#13& 4& $\ket{\textrm{GHZ}(\pi/4)}_4$   &  2x2x2x2    &   0.3536     \\
\#14&  &                                 &  ~~{\footnotesize 3x3x3x2} & ~~{\footnotesize 0.3536}   \\ 
\#15&  &                                 &  ~~{\footnotesize 4x4x4x2} & ~~{\footnotesize 0.3536}   \\ 
\#16&  &                                 &  3x3x3x3    &   0.3408 \\ \cline{3-5}
\#17&  & $\ket{W}_4$                     &  2x2x2x2    &   0.5469 \\ 
\#18&  &                                 &  ~~{\footnotesize 3x3x3x2} & ~~{\footnotesize 0.5011}    \\ 
\#19&  &                                 &  ~~{\footnotesize 4x4x4x2} & ~~{\footnotesize 0.4923}    \\ 
\#20&  &                                 &  3x3x3x3    &   0.4900 \\ \cline{3-5}
\#21&  & $\ket{\Psi_4}$                  &  2x2x2x2    &   0.5303          \\
\#22&  &                                 &  ~~{\footnotesize 3x3x3x2} & ~~{\footnotesize 0.5012}   \\ 
\#23&  &                                 &  ~~{\footnotesize 4x4x4x2} & ~~{\footnotesize 0.4948}   \\ 
\#24&  &                                 &  3x3x3x3    &   0.4823          \\ \cline{3-5}
\#25&  & $\ket{D_4^{(2)}}$               &  2x2x2x2    &   0.4714          \\
\#26&  &                                 &  ~~{\footnotesize 3x3x3x2} & ~~{\footnotesize 0.4646}   \\ 
\#27&  &                                 &  ~~{\footnotesize 4x4x4x2} & ~~{\footnotesize 0.4630}    \\ 
\#28&  &                                 &  3x3x3x3    &   0.4393          \\ \cline{3-5}
\#29&  & $\ket{\textrm{Cluster}}_4$      &  2x2x2x2    &   0.5000          \\
\#30&  &                                 &  3x3x3x3    &   0.4472          \\ \hline 
\#31& 5& $\ket{\textrm{GHZ}(\pi/4)}_5$   &  2x2x2x2x2  &   0.2500          \\
\#32&  &                                 &  3x3x3x3x3  &   0.2280          \\  \cline{3-5}
\#33&  & $\ket{W}_5$                     &  2x2x2x2x2  &   0.4300          \\
\#34&  &                                 &  3x3x3x3x3  &   0.3462          \\ \cline{3-5}
\#33&  & $\ket{\textrm{Cluster}}_5$      &  2x2x2x2x2  &   0.3333          \\
\#34&  &                                 &  3x3x3x3x3  &   0.3333          \\ \hline 
\#35& 6& $\ket{\textrm{GHZ}(\pi/4)}_6$   &  2x2x2x2x2x2&   0.1768          \\  \cline{3-5}
\#36&  & $\ket{W}_6$                     &  2x2x2x2x2x2&   0.2927          \\ \cline{3-5}
\#37&  & $\ket{\Psi_6}$                  &  2x2x2x2x2x2&   0.3536          \\ \cline{3-5}
\#38&  & $\ket{D_6^{(3)}}$               &  2x2x2x2x2x2&   0.2827          \\ \hline 
\#39& 7& $\ket{\textrm{GHZ}(\pi/4)}_7$   &  2x2x2x2x2x2x2& 0.1250          \\ \cline{3-5}
\#40&  & $\ket{W}_7$                     &  2x2x2x2x2x2x2& 0.1975           \\ 
\hline \hline
\end{tabular}
\caption{The critical visibilities for various types of quantum states and experimental situations (number of settings). If $v>v_{crit}$, there does not exist any local realistic model describing quantum probabilities of experimental events.}
\label{tab-states}
\end{table}

{\em Observation 1.} An explicit form of tight Bell inequalities for probabilities is known only for the following experimental situations: 2x2 \cite{CH}, 3x3 \cite{SLIWA, SVOZIL}, 2x2x2 \cite{SLIWA, SVOZIL}, whereby $i$x$j$ means $i$($j$) alternative measurements on Alice's (Bob's) side.  In these cases the numerical method gives the same critical visibility as the analytical expressions. For states  (\#1 - \#5) (see Tab. \ref{tab-states}) we observe such a correspondence.

{\em Observation 2.} Local realistic correlations in the $N$-qubit experiments with two alternative measurement settings per side are bounded by WWWZB
inequalities \cite{WW,WZ,ZB}. In this case, the numerical method should give better or at least the same results. We obtain the same results for the following states: $\textrm{GHZ}(\pi/4)_N$ (\#1, \#2, \#13, \#31, \#35, \#39), $\Psi_4$ (\#21), $D_4^{(2)}$ (\#25), $\textrm{Cluster}_4$ (\#29), $\Psi_6$ (\#37), $D_6^{(3)}$ (\#38). Stronger constraints on local realism than predicted by the WWWZB inequalities are observed for the following states: 
\begin{itemize}
\item $\ket{\textrm{W}}_N$ (\#17, \#33, \#36, \#40). The corresponding critical visibilities obtained by means of WWWZB inequalities are equal to: $v^{crit}_{W_3} = 0.6565$, $v^{crit}_{W_4} = 0.6325$, $v^{crit}_{W_5} = 0.6202$, $v^{crit}_{W_6} = 0.6141$, and $v^{crit}_{W_7} = 0.6089$, and are higher than obtained by the numerical method by about 2\% (for $N=3$) to 208\% (for $N=7$). The values (\#5, \#17, \#33) were previously published in \cite{SEN}.

\item $\textrm{GHZ}(\pi/12)_N$ (\#3), $\textrm{GHZ}(\pi/180)_N$ (\#4). These states do not violate WWWZB inequalities at all \cite{SCARANI,ZBLW}. 

\end{itemize}
One of the simplest examples showing the advantage of Bell inequalities based on probabilities over correlation functions is a three qubit state, which is the product of the two qubit singlet state and white noise $|\psi^-\rangle_{ab} \langle \psi^-| \otimes \openone_c$. The state does not violate the WWWZB inequalities for three qubits, whereas the numerical method detects violation of local realism and gives the critical visibility equal to 0.7071. This number is equivalent to the visibility, which is necessary to violate the Clauser-Horne inequality by the $|\psi^-\rangle_{ab}$ state alone. The advantage comes from the fact that probability methods use the whole knowledge obtained from an experiment. 


{\em Observation 3.} Explicit forms of tight correlation function Bell inequalities \cite{n2} 
for three qubits and three settings per side were given in \cite{WIESZ}. The numerical method gives the same result in (\#2). In the case of (\#3, \#4), the inequalities of \cite{WIESZ} are not violated. The critical visibility for violation of the inequalities by the $\textrm{W}_3$ state (\#5) is equal to 0.6547 (about 2\% above the visibility predicted by the numerical method).

{\em Observation 4.} Explicit forms of tight correlation function Bell inequalities for many qubits ($N>3$) and many measurement settings ($m_i>2$) are still unknown. Tight Bell inequalities were constructed for some particular cases (e.g. $2^{N-1}{\rm x}2^{N-1}{\rm x}2^{N-2}\rm{x}\dots\rm{x}2$ \cite{LPZB}) only. There is also the family of the correlation Bell inequalities for many qubits and an arbitrary number of settings \cite{NLP}. Unfortunately, these inequalities are not tight. In this case, the numerical method gives the strongest known constraints on a local realistic description. 
\begin{itemize}
\item First, let us consider the 4x4x2-type inequality of \cite{LPZB}. It is violated by the ${\rm GHZ}(\pi/4)_3$ state with a critical visibility equal to 0.3536 which agrees with (\#2). However, for the ${\rm W}_3$ state the inequality is violated if the visibility is higher than 0.6547 \cite{LPZB}, whereas the numerical method reveals the critical visibility to be lower by about 4.9\% (\#7). 
\item One can also compare the numerical method with so called geometric inequalities \cite{NLP}. For instance, the critical visibility of the four qubit GHZ state to violate the three-setting geometrical inequality $v_{geom}^{GHZ_4}=0.3421$ is close to (but higher than) that expected by the numerical method (\#16). However, for the four-qubit W state the three-setting geometrical inequalities are violated with a critical visibility $v_{geom}^{W_4} =0.6843$, whereas the numerical method provides a drastically (about 28\%) better result (\#20). 
\end{itemize}

{\em Observation 5.} One can conjecture that the critical visibility for the three-qubit W state for an arbitrary large number of settings is equal to 6/10 (\#12).

{\em Observation 6.} In many cases, the critical visibility for impossibility of a local realistic description decreases with the number of settings. The highest visibility drop (19\%) is observed for the five-qubit W state (\#33, \#34), 10\% for the four-qubit W state (\#17, \#20), and less for ${\rm W}_3$ (\#5, \#10 - \#12), ${\rm GHZ}(\pi/4)_4$ (\#13, \#16), ${\rm GHZ}(\pi/4)_5$ (\#31, \#32), ${\rm Cluster}_4$ (\#29, \#30), $\Psi_4$ (\#21, \#24), and $D_4^{(2)}$ (\#25, \#28). Interestingly, for two- and three-qubit GHZ states we do not observe this effect. Acording to \cite{HUN} it could be that the effect appears with higher numbers of measuremaent settings. 

{\em Observation 7.} The minimal number of measurement settings determines a lower bound on the critical visibility. In (\#5 - \#10, \#13 - \#16, \#17 - \#20, \#21 - \#24, \#25 - \#28) the last observer has only two alternative measurement settings. Increasing the number of measurement settings at the other sides does not result in the decrease of the critical visibility below the value for three settings per each observer.

{\em Observation 8.} The $N$-qubit GHZ($\pi/4$) state reveals the strongest non-classical properties (the lowest critical visibility for an arbitrary number of $N \leq 8$). However, one can observe that the difference between the critical visibility for the GHZ state and the W state (or the cluster state) decreases with the number of qubits.  According to \cite{SEN}, we know that the W state leads to stronger nonclassicality than GHZ states for $N>10$. Finding an example for $N \leq 10$ is still an open problem. Due to computer power limitations, we were not able to recover the result of Sen {\it et. al} or find another one for $N \sim 10$.

\subsection{Bound entanglement}

We also analyze some classes of bound entangled states, namely,
\begin{itemize}
\item the three qubit Bennett state \cite{BENNETT}:
$$\rho_{\textrm{Bennett}} = \frac{1}{4}(\openone - \sum_{i=1}^4\ket{\phi_i}\langle\phi_i|),$$ where 
$\ket{\phi_1}=\ket{01+}, 
\ket{\phi_2}= \ket{1+0}, 
\ket{\phi_3}=\ket{+01}, 
\ket{\phi_4}=\ket{---}, 
|\pm\rangle = (\ket{0} \pm \ket{1})/\sqrt{2}$, 
\item the $N$-qubit D\"ur state \cite{DUR}:
$$\rho^{\textrm{D\"ur}}_N = \frac{1}{N+1} \left( |\phi \rangle \langle \phi| + \frac{1}{2} \sum_{k=1}^N (P_k + \tilde P_k) \right)$$ 
with $|\phi \rangle = \frac{1}{\sqrt{2}} \left[ |0 \rangle_1 ... | 0 \rangle_N + e^{i \alpha_N} |1 \rangle_1 ... | 1\rangle_N \right]$, and $P_k$ being a projector on the state $|0 \rangle_1 ... |1 \rangle_k ... |0 \rangle_N$ with ``1'' on the $k$th position ($\tilde P_k$ is obtained from $P_k$ after replacing ``0''s by ``1''s and vice versa),
\item the generalized $N$-qubit Smolin state \cite{AH}:
$$\rho^{\textrm{Smolin}}_{N} = \frac{1}{2^{N}}\left[\openone^{\otimes N}+(-1)^{N/2}\sum_{i=1}^{3}\sigma_{i}^{\otimes N}\right],$$ 
where $\sigma_i$ are the Pauli matrices.
\end{itemize}

The results are presented in Tab. \ref{tab-bound} and lead to the following observations.

\begin{table}
\begin{tabular}{l l l c c} \hline \hline
No.&$N$ & state & $m_1~ \textrm{x}~ m_2 ~\textrm{x}~ \cdots ~\textrm{x}~ m_N$ & $v_{crit}$  \\ \hline \hline
\#41&3& $\ket{\textrm{Bennett}}_3$  &     2x2x2 - 5x5x5            &    1.0000         \\ \hline
\#42&4& $\ket{\textrm{D\"ur}}_4$  &      2x2x2x2 - 3x3x3x3                 &    1.0000       \\ 
\#43& & $\ket{\textrm{Smolin}}_4$  &  2x2x2x2      &  0.7071          \\ 
\#44& &                            &  3x3x3x3      &  0.6986          \\ \hline
\#45&5& $\ket{\textrm{D\"ur}}_5$  &   2x2x2x2x2 -3x3x3x3x3  &     1.0000        \\  \hline
\#46&6& $\ket{\textrm{D\"ur}}_6$  &   2x2x2x2x2x2           &     1.0000       \\ 
\#47& & $\ket{\textrm{Smolin}}_6$  &  2x2x2x2x2x2           &     0.7071        \\  \hline
\#48&7& $\ket{\textrm{D\"ur}}_7$  &   2x2x2x2x2x2x2         &     1.0000        \\ 
\hline \hline
\end{tabular}
\caption{The critical visibilities for some bound entangled states and experimental situations (number of settings). If $v>v_{crit}$, there does not exist any local realistic model describing quantum probabilities of experimental events.}
\label{tab-bound}
\end{table}

{\em Observation 9.} Violation of local realism by the bound entangled Smolin state (for four and six qubits). The critical visibilities obtained by the numerical method (\#43, \#47) recover the results of \cite{AH}. If we go to three alternative settings per side, the critical visibility for the four qubit Smolin state decreases (\#44).

{\em Observation 10.} In \cite{DUR, KASZ3, NLP} the problem of the violation of Bell-type inequalities by the D\"ur state is considered. 
A simple experimental situation involves 7 qubits and 3 settings \cite{KASZ3}, 6 qubits and 5 settings  \cite{NLP}, or 8 qubits and 2 settings per side \cite{DUR}.  We only verified the case of 7 qubits (\#48). Using the numerical method we can not find a violation of a local realistic model for two settings per side. It suggests that the requirement of three settings is reasonable. We also partially verified the cases of 4 (\#42), 5 (\#45) and 6 (\#46) qubits. We can not find a local realistic model for two measurement settings per side (and three setting in the case of 4 and 5 qubits). The problem of a violation of local realism by predictions of the four and five qubit D\"ur state, and the six qubit D\"ur state with the number of measurement settings less than 5 is still an open question. 

Exemplarily, in App. \ref{LR-bound} we give a local realistic model for quantum probabilities in a Bell experiment for the four qubit D\"ur state. The observers choose between two measurement settings which are optimal for the case of the GHZ state.

\begin{figure}
\includegraphics[width=0.46\textwidth]{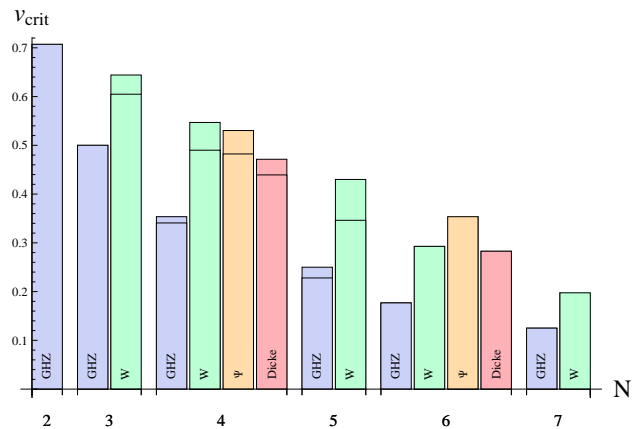}
\caption{(Color online) The bars represent for several families of quantum states the critical visibilities, which are necessary to falsify a local realistic description in two setting experiments. The additional line in some cases corresponds to the value of the critical visibility in a three setting experiment.}
\label{fig}
\end{figure}

\section{From detector clicks to density matrix}
\label{sectionIV}

In this section we use the numerical method for experimentally observed states. To this end, we need 
their density matrices. 
Methods that are used to obtain the density matrix from the measured 
data are called quantum state tomography. Different ways for quantum state
tomography have been developed (one of the first works was \cite{JAMES}; a variety of methods
is summarized in \cite{REHACEK}). From a decomposition of the density matrix into projectors,
one can easily express the density matrix in terms of probabilities for detecting a certain
coincidence. Relative frequencies obtained in a measurement are, however, subject to
Poissonian counting statistics. Because of these uncertainties, the deduced density matrices might be unphysical, e.g. one might find negative eigenvalues. Such raw data are not useful for the numerical method and lead
to the violation of local realism with visibility equal to 0, as spurious signaling-like effects may lurk in the raw data (due to experimental drifts). Fortunately, several different approaches for fitting a general
physical density matrix to the measured data have been developed. There are, however,
still discussions about the best method.

We apply the numerical method to the data obtained in the four-photon experiment reported in Refs.~\cite{EXP, IEEE, exp-nat}.
Spontaneous parametric down conversion in combination with linear optics was used to observe a variety of four-photon entangled states:
\begin{eqnarray}
|\psi(\gamma)\rangle &=& \frac{1}{2} \gamma (\ket{01}+ \ket{10})^{\otimes 2} \nonumber \\
&+& \sqrt{\frac{1-\gamma^2}{2}} (\ket{0011} + \ket{1100}).
\end{eqnarray}
We select three of these states, namely, $\ket{\psi(0)} \equiv \ket{GHZ(\pi/4)}_4$,  $\ket{\psi(1/3)} \equiv \ket{\Psi_4}$, $\ket{\psi(2/3)} \equiv \ket{D_4^{(2)}}$.  Moreover, two further states, namely: $\ket{GHZ(\pi/4)}_2$ and  $\ket{W}_3$ are obtained from the state $|\psi(2/3)\rangle$ after projective measurements. 
The experimental states were deduced from a full quantum state tomography with a maximum-likelihood estimation \cite{HRADIL} of the most likely density matrix to have led to the observed data. 

The \textsc{Steam-roller} was used to determine the critical parameter $v_{crit}^{exp}$ for the tomographically reconstructed states (see Tab. \ref{tab-exp}). If $v_{crit}^{exp}<1$, the experiment cannot be described by any local realistic model. 
The critical parameters $v_{crit}^{exp}$ can be compared with theoretical ones $v_{crit}^{theor}$ obtained for ideal states. The critical visibility in the experiment is always larger than the theoretical one. This is due to the non-ideal preparation of the states in the experiment, which essentially reduces the non-classical correlations. Further, the ratios $v_{crit}^{theor}/v_{crit}^{exp}$ are slightly greater than the experimental fidelity of a given state, e.g. for the Dicke state (\#57) $v_{crit}^{theor}/v_{crit}^{exp} = 0.86$, whereas $F^{exp}_{Dicke} = 0.83$.
This can be explained by the fact, that the noise observed in the experiment contains nonclassical correlations \cite{LASKOWSKI, MUNICH}.

\begin{table}
\begin{tabular}{l l l c c c} \hline \hline
No.& $N$ & State & $m_1~ \textrm{x}~ m_2 ~\textrm{x}~ \cdots ~\textrm{x}~ m_N$ & $v_{crit}^{theor}$ & $v_{crit}^{exp}$ \\ \hline \hline
\#49&$2$ & $\ket{\textrm{GHZ}(\pi/4)}_2$ &  2x2    & 0.7071 & 0.7751 \\ 
\#50& &  &  3x3    & 0.7071 & 0.7737 \\ \hline
\#51&$3$ &  $\ket{W}_3$                &  2x2x2  & 0.6442 & 0.7534        \\ 
\#52&    &                             &  3x3x3  & 0.6048 & 0.7073 \\ \hline
\#53&$4$ & $\ket{\textrm{GHZ}(\pi/4)}_4$ & 2x2x2x2 & 0.3536 & 0.4365 \\
\#54&    &                             & 3x3x3x3 & 0.3408 & 0.4284 \\
\#55&    & $\ket{\Psi_4}$              & 2x2x2x2 & 0.5303 & 0.5748 \\
\#56&    &                             & 3x3x3x3 & 0.4823 & 0.5410 \\ 
\#57&   & $\ket{D}_4^{(2)}$           & 2x2x2x2 & 0.4714 & 0.5501 \\
\hline \hline
\end{tabular}
\caption{The critical visibilities for some states ({\em theor} - theoretical and {\em exp} - experimental) and number of measurement settings.}
\label{tab-exp}
\end{table}

\section{Closing remarks}

In this paper we used linear programming as a tool to study the non-classicality of quantum states. The detailed comparison of a number of multipartite entangled states demonstrates the power of this method even for modest computer equipment. Most of the conclusions were spelled out in the earlier sections. Here we want to stress that the overall message of the obtained data is that with more settings per observer one gets stronger violations of local realism. The strength of violation also increases with the number of qubits (see Fig. \ref{fig}).

We would like to draw the attention of the reader to the peculiarity of three-qubit GHZ states. It is well known that the threshold visibility for the GHZ correlations to violate the Mermin three qubit inequality is $0.5$. This inequality only considers highest order correlation functions. This threshold is very stubborn. As outlined previously, in our numerical approach we used not the correlation functions, but the full set of probabilities, for events of all orders. Still, even if one increases the number of settings from two per observer to five the stubborn threshold stays put at $0.5$, see also \cite{FOOTNOTE}. An interesting extension of the phenomenon are four-qubit GHZ states:
if one increases the number of settings from two to four for \emph{three} observers only, with the last one always using two settings, the threshold visibility is $0.3536$ for all such cases. The exact value seems to be $\frac{1}{2\sqrt{2}}$, as such is the analytic value for standard Bell inequalities (in the two settings per observer scenario). However in the 3x3x3x3 case, suddenly the critical visibility drops to $0.3408$. All this points to some so far unexplained resistance of three-particle correlations, even within a four-particle GHZ state, to reveal more non-classicality with an increasing number of settings. It should be an exciting task to find a reason for that behavior.

\section{Acknowledgments}

We acknowledge support from the EU program QAP (Contract No. 015848) and Q-ESSENCE (Contract No. 248095).
J.G., W.L., and M.\.Z. are supported by the MNiSW Grant no. N202 208538.
W.L. is supported by the Foundation for Polish Science (KOLUMB program). 
W.W. acknowledges support by QCCC of the ENB. 
N.K. acknowledges support from the Alexander 
von Humboldt Foundation. 
The collaboration is a part of a DAAD/MNiSW program.
M.\.Z. thanks Nicolas Gisin for discussions on the problem studied in section \ref{sectionIV}.

\appendix

\section{Local realistic model for the noisy $|GHZ(\pi/4)\rangle_2$ state}

The quantum probability function for the ${\rm GHZ}(\pi/4)_2$ state has the following form:
\begin{eqnarray}
&&P_{QM}(r_a, r_b | \theta_a, \phi_a; \theta_b, \phi_b) \nonumber \\
&&= \cos{(r_a \pi/ 4 + \theta_a)}\cos{(r_b \pi/ 4 + \theta_b)} \label{ghz2} \\
&&+ \sin{(r_a \pi/ 4 + \theta_a)}\sin{(r_b \pi/ 4 + \theta_b)} \cos{(\phi_a + \phi_b)} .\nonumber
\end{eqnarray}
Let us choose some arbitrary set of measurement angles: $\theta_i^j =0, \phi_1^1=0; \phi_1^2=\pi/2; \phi_2^1=\pi/4; \phi_2^2=3\pi/4$ and put (\ref{ghz2}) to (\ref{marginals}). Then we get the set of constraints, under which the visibility function $v$ should be maximized:
\begin{eqnarray}
p_0 + p_1  + p_5 + p_4 &=& q_- \nonumber \\
p_2 + p_3 + p_7  + p_6 &=& q_+\nonumber \\
p_4 + p_5  + p_{13} + p_{12} &=&q_- \nonumber \\
p_6 + p_7  + p_{14} + p_{15} &=&q_+ \nonumber \\
p_0 + p_2  + p_8 + p_{10} &=& q_+\nonumber \\
p_1 + p_3 + p_9  + p_{11} &=& q_-\nonumber \\
p_4 + p_6  + p_{12} + p_{14} &=& q_-\nonumber \\
p_5 + p_7  + p_{13} + p_{15} &=& q_+\nonumber \\
p_8 + p_9  + p_{13} + p_{12} &=& q_+\nonumber \\
p_{10} + p_{11}  + p_{14} + p_{15} &=& q_-\nonumber \\
p_0 + p_2  + p_6 + p_4 &=& q_+\nonumber \\
p_1 + p_3 + p_7  + p_5 &=& q_-\nonumber \\
p_8 + p_{10} + p_{12} + p_{14} &=& q_-\nonumber \\
p_9 + p_{11}  + p_{13} + p_{15} &=& q_+\nonumber \\
p_0 + p_1 + p_9  + p_8 &=& q_+\nonumber \\
p_2 + p_3 + p_{11}  + p_{10} &=& q_-
\end{eqnarray}
with $q_{\pm} =(1 \pm  v / \sqrt{2})/4$, $p_i \leq 1$ and $\sum_i p_i =1$. For such a problem the numerical procedure 
gives the highest possible visibility $v_{crit} = 0.707107$, for which a local
realistic model of (\ref{ghz2}) exists. For such a value of $v=v_{crit}$, the local realistic model has the following form:
\begin{equation}
\begin{array}{l l l l}
p_0  = 0,&
p_1  = 1/8,&
p_2  = 0,&
p_3 = 1/8,\\
p_4  = 1/8,&
p_5  = 1/8,&
p_6  = 0,&
p_7  = 0,\\
p_8  = 0,&
p_9  = 0,&
p_{10}  = 1/8,&
p_{11}  = 1/8,\\
p_{12}  = 1/8,&
p_{13}  = 0,&
p_{14}  = 1/8,&
p_{15} = 0.
\end{array}
\end{equation}
Note that in this example, the measurement was chosen arbitrarily (but in an optimal way). In a standard procedure, there is an additional optimization over measurement angles. 

\section{Local realistic model for a bound entangled state \label{LR-bound}}

The local realistic model for the quantum probabilities for the four-qubit D\"ur state is presented. The measurement angles are chosen in the optimal way for the ${\rm GHZ}(\pi/4)_4$ state, namely, $\theta_i^j =0 (i=1,2,3,4; j=1,2), \phi_1^1=0, \phi_1^2=\pi/2, \phi_2^1=\pi/4, \phi_2^2=3\pi/4, \phi_3^1=\pi/4, \phi_3^2=3\pi/4, \phi_4^1=\pi/4, \phi_4^2=3\pi/4$. Any other choice of measurements does not have an impact for violation of local realism. The model has the following form:
\begin{itemize}
\item $p_{0}= p_{15}=  p_{21}=  p_{26}=  p_{38}=  p_{41}=  p_{51}= 
p_{60}= p_{67}=  p_{76}=  p_{112}= p_{127}= p_{150} = p_{153}= 
p_{165}= p_{170}= p_{214}= 0.0404029$,

\item $p_{85}=  p_{90}=  p_{102}= p_{105}= p_{128}= p_{143}= p_{179}= 
p_{188}= p_{195}= p_{200}= p_{204}= p_{217}=  p_{229}=  p_{235}= 
p_{240}=  p_{248}=  p_{255}=  0.00379126$,

\item $p_{197}= p_{219}= p_{226}= p_{238}= p_{245}= 0.0366117$,

\item $p_{216}=  p_{233}=  0.0328204$.
\end{itemize}
All other probabilities vanish.


\begin{thebibliography}{99}

\bibitem{Ekert91}
A. K. Ekert, Phys. Rev. Lett. {\bf 67}, 661 (1991).

\bibitem{ZukEtAll1998ActaPhysPolA}
M.\.Zukowski, A. Zeilinger , M.A. Horne and H. Weinfurter, Acta Phys. Pol. {\bf 93}, 187 (1998).

\bibitem{Burhman1997}
H. Buhrman, W. van Dam, P. Hoyer, A. Tapp, Phys. Rev. A, {\bf 60}, 2737 (1999).

\bibitem{Galvao2002}
Ernesto F. Galvao, Phys. Rev. A {\bf 65}, 012318 (2001).

\bibitem{BrukZuk2004}
C. Brukner, M. Zukowski, J.-W. Pan, A. Zeilinger, Phys. Rev. Lett. {\bf 92}, 127901 (2004).


\bibitem{WW} R. F. Werner and M. W. Wolf, Phys. Rev. A {\bf 64}, 32112 (2001).

\bibitem{WZ} H. Weinfurter and M. \.Zukowski,Phys. Rev. A {\bf 64}, 010102(R) (2001).


\bibitem{ZukowskiBaturo98}
M. \.Zukowski, D. Kaszlikowski, A. Baturo, J.-A. Larsson, arXiv:quant-ph/9910058.

\bibitem{KASZLIKOWSKI-SR} 
D. Kaszlikowski, P. Gnaci\'nski, M. \.Zukowski, W. Miklaszewski and A. Zeilinger,
Phys. Rev. Lett. {\bf 85}, 4418 (2000).

\bibitem{ZB} M. \.Zukowski, {\v C}. Brukner, Phys. Rev. Lett. {\bf 88}, 210401 (2002).

\bibitem{n1}
The highest order correlation function describes correlations only between all subsystems.

\bibitem{BELL}
J. S. Bell, Pysics {\bf 1}, 195 (1964).


\bibitem{simplex}
G. B. Dantzig, Linear Programming and Extensions, Princeton University Press, Princeton, NJ (1963).

\bibitem{GLPK}
GNU Linear Programming Kit, Version 4.31, http://www.gnu.org/software/glpk/

\bibitem{dsm}
J. A. Nelder, R. Mead, Computer Journal {\bf 7}, 308 (1965).

\bibitem{scipy}
SciPy Python scientific computing package, Version 0.7.0, http://www.scipy.org/

\bibitem{GHZ} D. M. Greenberger, M. A. Horne, and A. Zeilinger, in
{\em Bell's Theorem, Quantum Theory, and Conceptions of the
Universe}, edited by M. Kafatos, Kluwer Academic, Dordrecht, 69 (1989).

\bibitem{SCARANI}
V. Scarani, N. Gisin, J. Phys. A: Math. Gen. {\bf 34}, 6043 (2001).

\bibitem{W}
W. D\"ur, G. Vidal, J. I. Cirac, Phys. Rev. A {\bf 62}, 062314 (2000).

\bibitem{CABELLO} A. Cabello, Phys. Rev. A {\bf 68}, 012304 (2003).

\bibitem{PSI6}
M. Radmark, M. \.Zukowski, M. Bourennane, Phys. Rev. Lett. {\bf 103}, 150501 (2009).

\bibitem{DICKE} R. H. Dicke, Phys. Rev. {\bf 93}, 99 (1954).

\bibitem{CLUSTER} H. J. Briegel and R. Raussendorf, Phys. Rev. Lett. {\bf 86}, 910 (2001).

\bibitem{CH} J. F. Clauser and M. A. Horne, Phys. Rev. D {\bf 10}, 526 (1974).

\bibitem{SVOZIL} I. Pitowsky and K. Svozil, Phys. Rev. A {\bf 64}, 014102 (2001).

\bibitem{SLIWA} C. \'Sliwa, Phys. Lett. A {\bf 165}, 3 (2003).

\bibitem{SEN}
A. Sen(De), U. Sen, M. Wie\'sniak, D. Kaszlikowski, and M. \.Zukowski, Phys. Rev. A {\bf 68}, 062306 (2003). 

\bibitem{ZBLW}
M. \.Zukowski, {\v C}. Brukner, W. Laskowski, M. Wie\'sniak, Phys. Rev. Lett. {\bf 88}, 210402 (2002).

\bibitem{n2}
Tight Bell inequalities define the half-spaces in which is the correlation (local realistic) polytope, which contain a face of it in their border hyperplane.

\bibitem{WIESZ} M. Wie\'sniak, Marek \.Zukowski, Phys. Rev. A {\bf 76} 012110 (2007).

\bibitem{LPZB} W. Laskowski, T. Paterek, M. \.Zukowski, {\v C}. Brukner, Phys. Rev. Lett. {\bf 93}, 200401 (2004).

\bibitem{NLP} K. Nagata, W. Laskowski, T. Paterek, Phys. Rev. A {\bf 74}, 062109 (2006).

\bibitem{HUN} T. V\'ertesi, Phys. Rev. A {\bf 78}, 032112 (2008).

\bibitem{BENNETT}
C.H. Bennett, D.P. DiVincenzo, T. Mor, P.W. Shor, J.A. Smolin, B.M. Terhal, Phys.Rev.Lett. {\bf 82} 5385 (1999).

\bibitem{DUR} W. D\"ur, Phys. Rev. Lett. {\bf 87}, 230402 (2001).

\bibitem{AH} R. Augusiak, P. Horodecki, Phys. Rev. A {\bf 73}, 012318 (2006).

\bibitem{KASZ3}
D. Kaszlikowski, L. C. Kwek, J. Chen, and C. H. Oh, Phys. Rev. A {\bf 66}, 52309 (2002).

\bibitem{JAMES}
D.F.V. James, P.G. Kwiat, W.J. Munro and A.G. White, Phys. Rev. A {\bf 64}, 052312 (2001).

\bibitem{REHACEK}
J. Rehacek and M.G.A. Paris: Lecture Notes in Physics: Quantum State Estimation, Springer, 2004.

\bibitem{EXP}
W. Wieczorek, Ch. Schmid, N. Kiesel, R. Pohlner, O. G\"uhne, and H. Weinfurter, Phys. Rev. Lett. {\bf 101}, 010503 (2008).

\bibitem{IEEE}
W. Wieczorek, N. Kiesel, C. Schmid, W. Laskowski, M. Zukowski and H. Weinfurter,
IEEE J. Sel. Top. Quant. Electron. {\bf 15}, 1704 (2009).

\bibitem{exp-nat}
M. Aspelmeyer and J. Eisert, Nature {\bf 455} 180 (2008).

\bibitem{HRADIL}
Z. Hradil, Phys. Rev. A {\bf 55}, R1561 (1997).

\bibitem{LASKOWSKI}
W. Laskowski, M. Wie\'sniak, M. \.Zukowski, M. Bourennane, H. Weinfurter, J. Phys. B {\bf 42}, 114004 (2009).

\bibitem{MUNICH}
W. Wieczorek, N. Kiesel, C. Schmid, and H. Weinfurter, Phys. Rev. A {\bf 79}, 022311 (2009).

\bibitem{FOOTNOTE}
{Unfortunately, at the moment we cannot extend our investigations to more settings than 5. This is because of the memory requirements of the program and speed of the machine. Further, we cannot be $100\%$ certain that the algorithm reaches the global minimum. Still,  the  program is constructed in such a way that it never gives too low values for $v_{crit}$. Thus, we are always on the safe side with our claims of non-classicality.}

\end{thebibliography}
\end{document}